\documentclass[prd,twocolumn,aps,showpacs,nofootinbib,nobibnotes,superscriptaddress,preprintnumbers]{revtex4-1}
\usepackage{epsfig}
\usepackage{graphicx}
\usepackage{bm}
\usepackage[usenames, dvipsnames]{color}
\usepackage{dcolumn}   
\usepackage[spanish,english]{babel}
\usepackage{bm}     
\usepackage{bbm}       
\usepackage{amssymb}  
\usepackage{amsmath}
\usepackage{latexsym}
\usepackage{float}
\usepackage{ifthen}
\usepackage{caption,subfig}
\usepackage{enumerate}
\usepackage{url}
\usepackage{caption,subfig}
\usepackage{amsopn}
\usepackage{hyperref}

\usepackage{amsfonts}
\usepackage{multirow}
\usepackage{array}
\usepackage{booktabs}
\usepackage{rotating}

\usepackage{ulem}
\def\clap#1{\hbox to 0pt{\hss#1\hss}}

\def\bea{\begin{eqnarray}}
\def\eea{\end{eqnarray}}
\def\be{\begin{equation}}
\def\ee{\end{equation}}

\begin{document}

\title{Dark Energy in the Swampland II}

\author{Lavinia Heisenberg} \email{lavinia.heisenberg@eth-its.ethz.ch}
\affiliation{Institute for Theoretical Studies, ETH Zurich, 
 Clausiusstrasse 47, 8092 Zurich, Switzerland}
 
\author{Matthias Bartelmann} \email{bartelmann@uni-heidelberg.de}
\affiliation{Universit\"at Heidelberg, Zentrum f\"ur Astronomie, Institut f\"ur Theoretische Astrophysik, Germany}

\author{Robert Brandenberger} \email{rhb@hep.physics.mcgill.ca}
\affiliation{Physics Department, McGill University, Montreal, QC, H3A 2
T8, Canada}

\author{Alexandre Refregier} \email{alexandre.refregier@phys.ethz.ch}
\affiliation{Institute for Astronomy, Department of Physics, ETH Zurich, Wolfgang-Pauli-Strasse 27, 8093, Zurich, Switzerland}

\date{\today}

\begin{abstract}

In this recent Letter \cite{us}, we studied the implications of string Swampland conjectures for quintessence models of dark energy. A couple of days ago, a paper appeared \cite{Linde} which refers to our work but misrepresents what we have done. Here, we set the record straight.

\end{abstract}


\maketitle


We have recently pointed out \cite{us} that the string swampland conjectures \cite{swamp}, if true, provide important constraints on dark energy models.  The constraints apply to the field range of a scalar field $\phi$ described by an effective field theory, and to the slope of the potential $V$ of such fields. Specifically, we considered the consequence of the constraint
\begin{equation}
V'|/V \, > \, \lambda \, \sim \, \mathcal{O}(1) 
\end{equation}
(in Planck units), where the prime indicates the derivative with respect to the field. Specifically, we considered exponential potentials of the form 
\begin{equation}
V(\phi) \, \sim \, e^{- \lambda \phi} \, .
\end{equation}
We found that, at the 2-$\sigma$ level, current data already rule out values $\lambda > 0.9$, and we studied the predicted constraints from future observations such as Euclid.

Quintessence, a slowly rolling scalar field, has been a popular candidate for dark energy. In particular, negative exponential potentials are often considered (see e,g, \cite{Ratra, Wetterich} for early works).  In \cite{us}, we have considered the implications of the swampland conjecture on this class of dark energy models and find that, for values $\lambda > 0.9$, these models are in conflict with current observational data at the 2-$\sigma$ level. It is completely misleading to say that, as is implied in the abstract and main text of \cite{Linde}, that we are proposing string theory models of dark energy which are already ruled out by cosmological observations. 

Instead, the main goal of our paper was to study to what level the bound on $\lambda$ can be reduced by future data. We found that with Euclid data, the bound might improve to $\lambda < 0.4$ (at the three sigma level). Since the string swampland conjecture yields a lower bound on $\lambda$, whereas data yields an upper bound, the exciting prospect emerges that, provided the theory bound on $\lambda$ can be pinned down and turns out to be close to $\lambda = 1$, that all quintessence models with exponential potentials end up in the swampland. Such a result would require theorists to go back to the drawing board to explain dark energy.

In the Letter \cite{us}, we investigated the consequences of the Swampland criteria on the existing dark energy models and we made the following three statements:
\begin{itemize} 
 \item Using the 1- and  2-$\sigma$ contours of Fig. 21 of \cite{Scolnic:2017caz} on the Chevallier-Polarski-Linder (CPL) parametrisation \cite{Chevallier:2000qy} and directly translating them to an upper bound on the reconstructed equation of state, we find that $\lambda\le0.9$ at 2-$\sigma$ level. Since \cite{Scolnic:2017caz} does not provide the 3-$\sigma$ contour, giving the 3-$\sigma$ bound $\lambda\lesssim1.3$ required us (as it would require anybody else) to extrapolate from the 1- and 2-$\sigma$ contours. This can of course only be taken as an approximation to the inverse Fisher matrix. Doing so, we estimated an upper bound at the 3-$\sigma$ level, which should not be confused with a rigorous determination. From the string theory point of view, the exact value of $\lambda$ is unknown, since only an order of magnitude $\mathcal{O}(1)$ can be given. We wanted to know whether current observations already imply a significant tension with the Swampland criteria, which would require an order-of-magnitude difference, which is clearly not the case -- independently of whether it is $\lambda\lesssim0.9$ at 2-$\sigma$ level \cite{us} or $\lambda\lesssim0.9$ at the 3-$\sigma$ level \cite{Linde}.
 
 At this point, since the authors of \cite{Linde} find our results ``mysterious'', a further comment may be in order. Our analytical procedure to estimate the contours of the inverse Fisher matrix allowed us to test that the bounds on $\lambda$ are very sensitive to the precise values of the best-fitting parameters as well as on the principal values and the orientation of the principal axes of the inverse Fisher matrix. The estimates on the upper bounds on $\lambda$ are thus substantially less secure than the procedure applied by \cite{Linde} may suggest.

 \item Improvement of the uncertainties with the Stage-4 experiments \cite{stage4} shrinks the allowed class of dark energy models by pushing $\lambda$ down to $\lambda\lesssim 0.4$ at the 3-$\sigma$ level. Since the exact inverse Fisher matrix of the Euclid forecasts \cite{euclidRedBook} is not available yet, we adopted the same orientation of the inverse Fisher matrix as in \cite{Scolnic:2017caz}. Again, our analytical method reveals that this limit is highly sensitive to the exact orientation of the principal axes of the inverse Fisher matrix, justifying the conservative estimate we found. Using their methods, \cite{Linde} claim to be able to push this bound to $\lambda\lesssim 0.15$. This may or may not be possible by different combinations of data sets. However, since none of the inverse Fisher matrices required for a truly rigorous statistical analysis is available yet, claiming such tighter bounds is premature. Yet, given the order-of-magnitude character of the swampland conjecture, the differences between the results of \cite{Linde} from ours do not justify any strong claims.
   
 \item As an outlook on possible future constraints, we also showed that there will be fundamental observational limitations on distinguishing a cosmological constant from a dark energy model with $\lambda\lesssim0.1$. Constraints of this sort might thus arrive at an order-of-magnitude difference from the swampland conjecture.
\end{itemize}

The analyses of the current bounds on $\lambda$ from current observations performed in our paper and in \cite{Linde} are based on different procedures, and should be taken as complementary results highlighting their uncertainties. We find that the 2-$\sigma$ lower bound is $\lambda\lesssim0.9$, whereas \cite{Linde} finds this at the 3-$\sigma$ level. The authors of \cite{Linde} strongly emphasize this difference and put a similarly strong emphasis on the difference between our bound of $\lambda\lesssim 0.4$ versus their bound of $\lambda\lesssim 0.15$ expected from the Stage-4 experiments. Our method shows, however, that even small changes in the assumed inverse Fisher matrix can substantially change the bounds on the allowed CPL parameters, thus weakening the bounds on $\lambda$.

As recognised by the authors of \cite{Linde}, our method is based on a Fisher-matrix analysis and an elliptical representation of the confidence contours. We even gave an analytical expression for the contours in Eq.\ 7 in \cite{us}. While the authors in \cite{Linde} claim that their results were more trustworthy, we argue based on our method that the bounds are sensitive to the exact form of the inverse Fisher matrix, on which neither we nor the authors of \cite{Linde} have precise information.

Our analytical expression of the 1- and  2-$\sigma$ contours of Fig. 21 of \cite{Scolnic:2017caz} offers a completely independent and alternative scrutiny of the underlying data, which reveals a source of uncertainty important for a reliable estimate of realistic and conservative bounds on $\lambda$. Yet, even though our results differ to some degree from those found in \cite{Agrawal:2018own,Scolnic:2017caz}, this difference is small enough to be neglected comparing to the order-of-magnitude swampland conjecture.

Thus, contrary to the claims made in \cite{Linde},
\begin{itemize} 
 \item we are not proposing any string theory model of dark energy which are already ruled out by cosmological observations, but rather deliver an upper bound on $\lambda$ to be compared with the lower bound imposed by the string swampland conjecture.
 \item The discrepancy on the absolute value of $\lambda$ is neither significant nor relevant, given that the different procedures applied and that the bounds found are within the same order of magnitude. Precise bounds on $\lambda$ are not reliably known. Instead of an unjustified emphasis on differences irrelevant for the arguments made, more effort should be put in understanding the reliability of the stringy arguments and their global implications for cosmology.
\end{itemize} 

In conclusion, we find that the misrepresentation of our work \cite{us} by \cite{Linde} needs to be set straight. Concerning the specific bounds on the constant $\lambda$, we have explained why the difference between our results and those of \cite{Linde} is neither ``mysterious'' nor relevant, and argued why our value is more realistic.

\section*{Acknowledgements}
We are grateful for useful discussions with ... . LH thanks financial support from Dr.~Max R\"ossler, the Walter Haefner Foundation and the ETH Zurich Foundation.  RB is supported in part by an NSERC Discovery Grant and by funds from the Canada Research Chair program.


\end{document}